\input{aipcheck}
\documentclass[ ]{aipproc}
\layoutstyle{8x11double}
\SetInternalRegister\hbadness{8000}

\usepackage{amsmath,amssymb}
\usepackage{graphicx}

\newcommand{\bsubeq}{\begin{subequations}}
\newcommand{\esubeq}{\end{subequations}}

\DeclareFontFamily{U}{rsf}{}
\DeclareFontShape{U}{rsf}{m}{n}{
  <5> <6> rsfs5 <7> <8> <9> rsfs7 <10-> rsfs10}{}
\DeclareMathAlphabet\Scr{U}{rsf}{m}{n}

\def\lapproxeq{\lower .7ex\hbox{$\;\stackrel{\textstyle
<}{\sim}\;$}}
\def\gapproxeq{\lower .7ex\hbox{$\;\stackrel{\textstyle
>}{\sim}\;$}}

\begin{document}

\allowdisplaybreaks{

\title
{Inflation With A Realistic SO(10) Model\footnote{Talk presented
by B. Kyae at PASCOS'05, Gyeongju, Korea (May 30 - June 4, 2005),
at PPP'05, YITP, Kyoto, Japan (June 20-24, 2005), and at COSMO'05,
Bonn, Germany (Aug. 28 - Sept. 1, 2005). This work is based on
Ref.~\cite{so10inf}.
%
}}



\author{Bumseok \sc{Kyae}}{
  address={Korea Institute for Advanced Study,
Cheongnyangni-dong, Dongdaemun-gu, Seoul 130-722, Korea},
email={bkyae@kias.re.kr},
}
\author{Qaisar \sc{Shafi}}{
address={Bartol Research Institute,
University of Delaware, Newark, DE 19716, USA},
email={shafi@bartol.udel.edu} }
 \copyrightyear {2005}

\begin{abstract}
We implement inflation within a realistic supersymmetric $SO(10)$
model in which the doublet-triplet splitting is realized through
the Dimopoulos-Wilczek mechanism, the MSSM $\mu$ problem is
resolved, and higgsino mediated dimension five nucleon decay is
heavily suppressed. The cosmologically unwanted topological
defects are inflated away, and from $\delta T/ T$, the $B-L$
breaking scale is estimated to be of order $10^{16}-10^{17}$ GeV.
Including supergravity corrections, the scalar spectral index $n_s
= 0.99\pm 0.01$, with $|dn_s/ d{\rm ln}k| \lapproxeq 10^{-3}$.
\end{abstract}

\date{\today}

\maketitle


In a class of supersymmetric (SUSY) models, inflation is
associated with spontaneous breaking of a gauge symmetry, such
that $\delta T/ T$ is proportional to $(M/M_{\rm Planck})^2$,
where $M$ denotes the symmetry breaking scale and $M_{\rm Planck}$
($\equiv 1.2\times 10^{19}$ GeV) denotes the Planck
mass~\cite{hybrid,review}. Thus, from measurements of $\delta
T/T$, $M$ is estimated to be of order $10^{16}$
GeV~\cite{hybrid,ss}. The scalar spectral index $n_s$ in these
models is very close to unity in excellent agreement with recent
fits to the data~\cite{wmap}. A $U(1)$ $R$-symmetry plays an
essential role in the construction of these inflationary models.
These models possess another important property, namely with the
minimal K${\rm\ddot{a}}$hler potential, the supergravity (SUGRA)
corrections do not spoil the inflationary scenario~\cite{review},
which has been realized with a variety of attractive gauge groups
including $SU(3)_c\times SU(2)_L\times SU(2)_R\times U(1)_{B-L}$
($\equiv G_{LR}$)~\cite{LR}, $SU(4)_c\times SU(2)_L\times SU(2)_R$
($\equiv G_{422}$)~\cite{422}, $SU(5)\times U(1)$~\cite{flipped},
and $SU(5)$~\cite{su5}. We aim to implement inflation within a
realistic $SO(10)$ model~\cite{so10inf}.

$SO(10)$ has two attractive features, namely, it predicts the
existence of right handed neutrinos as well as the seesaw
mechanism. These two features are very helpful in understanding
neutrino oscillations and also in generating a baryon asymmetry
via leptogenesis. Furthermore, it seems easier to realize
doublet-triplet (DT) splitting without fine tuning in $SO(10)$
(say via the Dimopoulos-Wilczek mechanism~\cite{dw}) than in
$SU(5)$.

To implement $SO(10)$ inflation we would like to work with a
realistic model with the following properties: DT splitting is
realized without fine tuning, and the low energy theory coincides
with the minimal supersymmetric standard model (MSSM).  The MSSM
$\mu$ problem should also be resolved, and higgsino mediated
dimension five ($d=5$) nucleon decay should be adequately
suppressed.  Gauge boson mediated nucleon decay is still present
with a predicted nucleon lifetime of order $10^{34}-10^{36}$ yrs.
Finally, matter parity is unbroken, so that the LSP is stable and
makes up the dark matter in the universe.  To achieve natural DT
splitting and the MSSM at low energies with $SO(10)$, we will
follow Refs.~\cite{miniso10,dim5}, with suitable modifications
needed to make the scheme consistent with the desired inflationary
scenario, and also to avoid potential cosmological problems
(monopoles, moduli, etc). While doing this we would like to also
ensure that the SUGRA corrections also do not disrupt the
inflationary scenario.

A minimal set of Higgs required to break $SO(10)$ to the MSSM
gauge group $SU(3)_c\times SU(2)_L\times U(1)_Y$ ($\equiv G_{SM}$)
is ${\bf 45}_H$, ${\bf 16}_H$, ${\bf\overline{16}}_H$.  A non-zero
vacuum expectation value (VEV) of ${\bf 45}_H$ along the $B-L$
($I_{3R}$) direction breaks $SO(10)$ to $G_{LR}$ ($SU(4)_c\times
SU(2)_L\times U(1)_R$) and produces magnetic monopoles. The ${\bf
16}_H$, ${\bf\overline{16}}_H$ VEVs break $SO(10)$ to $SU(5)$ and
induce masses for the right handed neutrinos via $d=5$ operators.
One of our goals is to make sure that the topological defects do
not pose cosmological difficulties.  Thus, it would be helpful if
during inflation $SO(10)$ is, for instance, broken to $G_{LR}$,
$SU(4)_c\times SU(2)_L\times U(1)_R$, or $G_{SM}$.

\begin{table}
\begin{tabular}{|c||cccccccccc|} \hline
 & $S$~ & $X$ & $X'$ & $Y$ & $P$ & $\overline{P}$ & $Q$ & $\overline{Q}$ &
  ${\bf 10}$ & ${\bf 10}_h$
 \\ \hline
$R$ & $1$ & $-1$ & $-1$ & $0$ & $0$ & $0$ & $-2$ & $2$ &
$1$ & $0$  \\
$A$ & $0$ & $-2/3$ & $-2/3$ & $-1/3$ & $-1/4$ & $1/4$ & $-1/2$ &
$1/2$ & $1/6$ & $0$
\\ \hline\hline
 & ${\bf 16}$ & ${\bf\overline{16}}$ & ${\bf 16}'$ & ${\bf \overline{16}}'$ &
${\bf 16}_H$ & ${\bf\overline{16}}_H$ & ~${\bf 16}_3$~ & ${\bf
45}$ & ${\bf 45}_H$ & ${\bf 45}_H'$
\\ \hline
$R$ & $1$ & $3$ & $2$ & $2$ & $0$ & $0$ & $1/2$ & $1$ & $0$ & $-1$

\\
$A$ & $1/2$ & $~2/3~$ & $2/3$ & $2/3$ & $0$ & $0$ & ~$0$~ & $1/2$
& $-1/6$ & $-1/3$
\\ \hline
\end{tabular}
\caption{$U(1)$ $R$ and $U(1)_A$ charge assignments for the
superfields.}
\end{table}
To implement DT splitting without fine tuning and eliminate $d=5$
proton decay, and to recover the MSSM at low energies with the
$\mu$ problem resolved, we need an additional ${\bf 45}$-plet
(${\bf 45}_H'$), two additional ${\bf 16}+{\bf \overline{16}}$
pairs, two ${\bf 10}$-plets (${\bf 10}_h$ and ${\bf 10}$), and
several singlets~\cite{miniso10,dim5}. One more ${\bf 45}$-plet is
also required by $U(1)$ $R$-symmetry. This symmetry, among other
things, plays an essential role in realizing inflation, and its
$Z_2$ subgroup coincides with the MSSM matter parity. The $SO(10)$
singlet superfields are denoted as $S$, $X$, $X'$, $Y$, $P$,
$\overline{P}$, $Q$, and $\overline{Q}$, whose roles will be
described below. TABLE 1 displays the quantum numbers (under the
global $U(1)$ $R$ and $U(1)_A$ symmetries) of all the Higgs sector
superfields and the third family matter field (${\bf 16}_3$). We
will take the same notation for the superfields and their scalar
components.

To break $SO(10)$ to $G_{LR}$, consider the superpotential,
\begin{eqnarray} \label{adj}
&&~~~W_{45}= \frac{\alpha}{6M_*} X^{(')}Y{\rm Tr}\left({\bf
45}^2\right) -\frac{\beta}{6} Y{\rm Tr}\left({\bf 45}{\bf
45}_H\right)
\\
&& +\frac{\gamma_1}{36M_*}{\rm Tr}\left({\bf 45}{\bf 45}_H\right)
{\rm Tr}\left({\bf 45}_H^2\right) +\frac{\gamma_2}{6M_*}{\rm
Tr}\left({\bf 45}{\bf 45}_H^3\right) \nonumber ~,
\end{eqnarray}
where $\alpha$, $\beta$, $\gamma_{1,2}$ are dimensionless
parameters, and $M_*$ $(\sim 10^{18}$ GeV) denotes the cutoff
scale. As will be explained, $X$, $X'$, and $Y$ can develop
non-zero VEVs, $\langle X\rangle \sim \langle X'\rangle \sim
\langle Y\rangle \sim 10^{16}$ GeV. Due to non-zero $\langle
Y\rangle$, ${\bf 45}_H$ can also obtain a VEV in the $B-L$
direction from the $\beta$ and $\gamma_{1,2}$ terms of
Eq.~(\ref{adj}),
\begin{eqnarray} \label{vev2}
\langle {\bf 45}_{H}\rangle = {\rm diag.}(v,v,v;0,0)\otimes
i\sigma_2 ~
\end{eqnarray}
and $\langle {\bf 45}\rangle =0$, where
$v=\sqrt{\frac{\beta}{\gamma}\langle Y\rangle M_*}\equiv M_{GUT}$
($\approx 3\times 10^{16}$ GeV), with $\gamma\equiv
\gamma_1+\gamma_2$. The upper-left $3\times 3$ block corresponds
to $SU(3)_c$ and the lower-right $2\times 2$ block to $SU(2)_L$ of
the $G_{SM}$. Hence, the $SO(10)$ gauge symmetry is broken to
$G_{LR}$. Note that from the `$\alpha$ term,' the ${\bf 45}$
multiplet becomes superheavy. It acquires a VEV of order
$(m_{3/2}M_{GUT})/M_*$ after SUSY breaking, where $m_{3/2}$
($\sim$ TeV) denotes the scale of the soft parameters.

The next step in the breaking to the MSSM gauge group $G_{SM}$
($=G_{LR}\cap SU(5)$) is achieved with the following
superpotential,
\begin{eqnarray}
W_{16}=S\left[\kappa {\bf 16}_H{\bf\overline{16}}_H
-\frac{\rho}{M_*^2}({\bf 16}_H{\bf\overline{16}}_H)^2+\lambda {\bf
10}_h^2 - \kappa M_{B-L}^2\right] \nonumber
\end{eqnarray}
\noindent $+{\bf 16}\left[\frac{\lambda_1}{M_*}{\bf
45}_HY-\frac{\lambda_2}{M_*}P^2 \right]{\bf\overline{16}}_H
+{\bf\overline{16}}\left[\frac{\lambda_3}{M_*}{\bf
45}_HQ-\frac{\lambda_4}{M_*} {\bf 45}_H^{'2}\right]{\bf 16}_H $
\begin{align} \label{spinor}
+ {\bf 16}'\left[\frac{\lambda_5}{M_*}{\bf 45}_H'Y-\lambda_6X
\right]{\bf\overline{16}}_H+{\bf\overline{16}}'\left[\frac{\lambda_7}{M_*}{\bf
45}_H'Y-\lambda_8X'\right]{\bf 16}_H
 ~,
\end{align}
where $\rho$ is a dimensionless coupling constant. The
dimensionful parameter $M_{B-L}$, as determined from inflation
($\delta T/T$), turns out to be of order $10^{16}-10^{17}$ GeV.
From the $\kappa$ and $\rho$ terms, and the ``D-term'' potential,
${\bf 16}_H$ and ${\bf \overline{16}}_H$ develop VEVs of order
$M_{B-L}$, breaking $SO(10)$ to $SU(5)$,
\begin{eqnarray} \label{vev3}
|\langle{\bf 16}_H\rangle|^2
=|\langle{\bf\overline{16}}_H\rangle|^2=
\frac{M_{B-L}^2}{2\zeta}\left[1-\sqrt{1-4\zeta}\right] ~,
\end{eqnarray}
where $\zeta\equiv\rho M_{B-L}^2/(\kappa M_*^2)$~\cite{422}, while
$\langle S\rangle =\langle{\bf 10}_h\rangle=0$ upto corrections of
$O(m_{3/2})$ by including soft SUSY breaking terms in the scalar
potential~\cite{LR}.  Together with Eq.~(\ref{vev2}), the $SO(10)$
is broken to the $G_{SM}$.  The MSSM Higgs doublets arise from
${\bf 10}_h$. With $\langle S\rangle\approx -m_{3/2}/\kappa$, the
$\mu$ term from Eq.~(\ref{spinor}) is of order
$(\lambda/\kappa)m_{3/2}\sim$ TeV.\footnote{ From $y_{\mu}{\bf
10}{\bf 10}_h\langle {\bf 16}_H{\bf 45}_H{\bf
\overline{16}}_H\rangle/M_*^2$, the doublets in ${\bf 10}_h$
obtains a ``seesaw mass'' $y_\mu^2(\langle {\bf 16}_H{\bf
45}_H{\bf \overline{16}}_H\rangle)^2/(M_*^4\langle {\bf
45}_H'\rangle)\sim$ TeV with $y_\mu \sim 10^{-3}$, which modifies
the $\mu$ parameter at low energies.} Similarly the soft term
$B\mu$ ($\approx -2(\lambda/\kappa)m_{3/2}^2$) is
generated~\cite{LR}.

When $SO(10)$ breaks to $G_{SM}$ by an adjoint and a vector-like
pair of spinorial Higgs, the superfields associated with $[\{{\bf
(3,2)}_{1/6}$, ${\bf (\overline{3},1)}_{-2/3}$, ${\bf
(\overline{3},1)}_{1/3}$, ${\bf (1,\overline{2})}_{-1/2}\}+{\rm
h.c.}]$ from the ${\bf 45}_H$ and ${\bf
16}_H$-${\bf\overline{16}}_H$ turn out to be pseudo-goldstone
modes~\cite{miniso10}. Such extra light multiplets would spoil the
unification of the MSSM gauge couplings, and therefore must be
eliminated.  The simplest way to remove them from the low energy
spectrum is to introduce couplings such as ${\bf 16}_H{\bf
45}_H{\bf\overline{16}}_H$. However, it destabilizes the form of
$\langle {\bf 45}_H\rangle$ given in Eq.~(\ref{vev2}), in such a
way that at the SUSY minimum, $v=0$ is required. It was shown in
Ref.~\cite{miniso10} that with the `$\lambda_i$' couplings
($i=1,2,3,4$) and an additional ${\bf 16}$-${\bf\overline{16}}$
pair in Eq.~(\ref{spinor}), the unwanted pseudo-goldstone modes
all become superheavy, keeping intact the form of Eq.~(\ref{vev2})
at the SUSY minimum.

From the ``F-flat conditions'' with ${\bf 16}_H$ and
${\bf\overline{16}}_H$ acquiring non-zero VEVs, one finds
\begin{eqnarray} \label{vev4}
\langle {\bf 45}_H\rangle\langle Y\rangle
=\frac{\lambda_2}{\lambda_1}\langle P^2\rangle ~,~~~\langle {\bf
45}_H\rangle\langle Q\rangle =\frac{\lambda_4}{\lambda_3}{\rm
Tr}\langle {\bf 45}_H'\rangle^2 ~.
\end{eqnarray}
Thus, if $P$ and $Q$ develop VEVs, $\langle{\bf 45}_H\rangle$,
$\langle {\bf 45}_H'\rangle$, and $\langle Y\rangle$ should also
appear. We will soon explain how $\langle P\rangle$ and $\langle
Q\rangle$ arise. Since $\langle Y\rangle$ is related to $\langle
{\bf 45}_H\rangle$ via Eq.~(\ref{vev2}), both are uniquely
determined. We assume that $\langle {\bf 45}_H'\rangle$ points in
the $I_{3R}$ direction,
\begin{eqnarray}
\langle {\bf 45}_{H}'\rangle = {\rm diag.}(0,0,0;v',v')\otimes
i\sigma_2 ~.
\end{eqnarray}
Recall that $\langle {\bf 45}_H'\rangle$ is employed to suppress
higgsino mediated $d=5$ nucleon decay~\cite{dim5}. Similarly, due
to the presence of the `$\lambda_i$' ($i=5,6,7,8$) couplings in
Eq.(\ref{spinor}), the low energy spectrum is protected even with
the ${\bf 45}_H'$ present~\cite{dim5}. With non-zero VEVs for
${\bf 45}_H'$ and $Y$, $X$ and $X'$ slide to the values satisfying
\begin{eqnarray} \label{vev5}
\frac{\lambda_{5,7}}{M_*}\langle {\bf 45}_H'\rangle\langle
Y\rangle-\lambda_{6,8}\langle X^{(')}\rangle=0 ~,
\end{eqnarray}
with $|\langle {\bf 16}'\rangle|=|\langle {\bf
\overline{16}}\rangle|\sim O(m_{3/2})$. In order to guarantee the
`$\lambda_i$' couplings in Eq.~(\ref{spinor}) and to forbid ${\bf
16}_H{\bf 45}_H{\bf\overline{16}}_H$,  the $U(1)$ symmetries in
TABLE 1 are essential.

To obtain non-vanishing VEVs for $P$ and $Q$, one could consider
the following superpotential,
\begin{align} \label{PQ}
W_{PQ}= S\left[\kappa_1P\overline{P}+\kappa_2Q\overline{Q}\right]
-\frac{S}{M_*^2}\left[\rho_1(P\overline{P})^2+\rho_2(Q\overline{Q})^2\right]~,
\end{align}
such that
$\langle P\overline{P}\rangle=(\kappa_1/\rho_1)M_*^2\sim \langle
Q\overline{Q}\rangle=(\kappa_2/\rho_2)M_*^2\sim M_{GUT}^2$.
The $\lambda_{2,3}$ terms in Eq.~(\ref{spinor}) just determine
$\langle {\bf 45}_H\rangle$, $\langle Y\rangle$, and $\langle{\bf
45}_H'\rangle$. With the inclusion of soft SUSY breaking terms,
$\langle P\rangle$, $\langle\overline{P}\rangle$, $\langle
Q\rangle$, and $\langle\overline{Q}\rangle$ would be completely
fixed. To avoid potential cosmological problems associated with
moduli fields, we assume that the VEVs satisfy the constraints
$\langle P\rangle=\langle\overline{P}\rangle$ and $\langle
Q\rangle=\langle\overline{Q}\rangle$. This could be made plausible
by assuming universal soft scalar masses, and that the SUSY
breaking ``A-terms" asymmetric under
$P\leftrightarrow\overline{P}$ and $Q\leftrightarrow\overline{Q}$
are small enough.\footnote{In gravity mediated SUSY breaking
scenario with the minimal K${\rm\ddot{a}}$hler potential, at the
minimum the ``A-terms'' corresponding to $\lambda_{k}$
($k=1,2,3,4$) are cancelled by each other with the VEVs in
Eq.~(\ref{vev4}).  The other soft terms are symmetric under
$P\leftrightarrow\overline{P}$ and $Q\leftrightarrow\overline{Q}$.
In gauge mediation, ``A-terms'' are generally suppressed.} Since
the fields that couple to $P$, $\overline{P}$, $Q$ and
$\overline{Q}$ are all superheavy, the soft parameters would be
radiatively stable at low energies. Thus, at the minimum of the
scalar potential, we have four mass eigen states,
$(P\pm\overline{P})/\sqrt{2}$ ($\equiv P_{\pm}$) and
$(Q\pm\overline{Q})/\sqrt{2}$ ($\equiv Q_{\pm}$). While $P_+$ and
$Q_+$ obtain superheavy masses and large VEVs of order $M_{GUT}$,
$P_-$ and $Q_-$ remain light ($\sim m_{3/2}$) with vanishing VEVs.

With $\langle{\bf 45}_H\rangle$ in Eq.~(\ref{vev2}), the ``DT
splitting problem'' can be resolved ~\cite{dw}. Consider the
superpotential:
\begin{eqnarray} \label{vec}
W_{10}=y_1{\bf 10}{\bf 45}_H'{\bf 10}+y_2{\bf 10}{\bf 45}_H{\bf
10}_h ~.
\end{eqnarray}
From the $y_1$ term, only the doublets contained in ${\bf 10}$
become superheavy~\cite{dim5}, and from the $y_2$ term only the
color triplet fields included in ${\bf 10}$ and ${\bf 10}_h$
acquire superheavy masses~\cite{dw,miniso10,dim5}. Since the two
color triplets contained in ${\bf 10}_h$ do not couple at all in
Eq.~(\ref{vec}), $d=5$ nucleon decay is eliminated in the SUSY
limit~\cite{dim5}. Note that operators such as ${\bf 10}{\bf
10}_h$, ${\bf 10}_h^2$, $[{\bf 10}{\bf 10}_h]{\rm Tr}({\bf
45}_H^2)$ and so on are allowed by $SO(10)$ and, unless forbidden,
would destroy the gauge hierarchy. The $U(1)$ symmetries in TABLE
1 are once again crucial in achieving this.

Although the superpotential coupling $\langle S\rangle{\bf
10}_h^2$ induces higgsino mediated $d=5$ nucleon decay, there is a
suppression factor of $m_{3/2}/M_{GUT}$. Thus, we expect that
nucleon decay is dominated by the exchange of the superheavy gauge
bosons with an estimated lifetime $\tau_p\rightarrow e^+ \pi^0$ of
order $10^{34}-10^{36}$ yrs. Note that we have assumed that $d=5$
operators such as ${\bf 16}_i{\bf 16}_j{\bf 16}_k{\bf 16}_l$,
${\bf 16}_i{\bf 16}_j{\bf 16}_k{\bf 16}_H$ and so on, where the
subscripts are family indices of the matter, are adequately
suppressed by assigning suitable $R$ and $A$ charges to these
matter fields.

Consider next the superpotential couplings involving the third
generation matter superfields,
\begin{eqnarray} \label{matter}
W_m=y_3{\bf 16}_3{\bf 16}_3{\bf 10}_h +\frac{y_{\nu}}{M_*} {\bf
16}_3{\bf 16}_3{\bf\overline{16}}_H{\bf\overline{16}}_H ~.
\end{eqnarray}
The first term yields Yukawa unification so that the MSSM
parameter ${\rm tan}\beta\approx m_t/m_b$.  From the $y_\nu$ term,
the right handed neutrino masses are $\lapproxeq y_\nu
M_{B-L}^2/M_*\sim 10^{14}$ GeV. Right handed neutrino masses of
order $10^{14}$ GeV and smaller can yield a mass spectrum for the
light neutrinos through the seesaw mechanism, that is suitable for
neutrino oscillations.  The role of `matter parity' is played by
the unbroken $Z_2$ subgroup of the $U(1)$ $R$-symmetry~\cite{LR}.
Thus the LSP in our model is stable and contributes to the dark
matter in the universe.


Let us now discuss how inflation is implemented in the model
described so far. The ``F-term'' scalar potential in SUGRA is
given by
\begin{eqnarray} \label{scalarpot}
V_F=e^{K/M_P^2}\bigg[\sum_{i,j}(K^{-1})^i_j(D_{\phi_i}W)(D_{\phi_j}W)^{*}
-3\frac{|W|^2}{M_P^2}\bigg] ~, \label{scalarpot2}
\end{eqnarray}
where $M_P$ ($\equiv M_{\rm Planck}/\sqrt{8\pi}=2.4\times 10^{18}$
GeV) denotes the reduced Planck mass. $K$ and $W$ are the
K${\rm\ddot{a}}$hler potential and the superpotential,
respectively. $(K^{-1})^i_j$ in Eq.~(\ref{scalarpot}) denotes the
inverse of $\partial^2 K/\partial\phi_i\partial\phi^*_j$. In our
case, $W$ is composed of Eqs.~(\ref{adj}), (\ref{spinor}),
(\ref{PQ}), (\ref{vec}), and (\ref{matter}). $D_{\phi_i}W$ is
defined as $\partial W/\partial\phi_i +(\partial
K/\partial\phi_i)(W/M_P^2)$. The K${\rm\ddot{a}}$hler potential
could be expanded as $K=|\phi_i|^2+c_4|\phi_i|^4/M_P^2+\cdots$.
Here, we consider the minimal case with $\partial^2
K/\partial\phi_i\partial\phi^*_j=\delta^i_j$. Indeed, higher order
terms in $K$ (with $c_4\lapproxeq 10^{-2}$) do not seriously
affect inflation~\cite{review}.

We employ the `shifted' hybrid inflationary scenario~\cite{422},
in which symmetries can be broken during inflation. An
inflationary scenario is realized in the early universe with the
scalar fields $S$, ${\bf 16}_H$, ${\bf \overline{16}}_H$, $P$,
$\overline{P}$, $Q$, and $\overline{Q}$ displaced from the present
values. We suppose that initially $|\langle S\rangle|^2 \gapproxeq
M_{B-L}^2[1/(4\zeta)-1]/2$ with $1/4<\zeta<1/7.2$~\cite{422}, and
$\langle{\bf 16}_H\rangle$, $\langle{\bf\overline{16}}_H\rangle$,
$\langle P\rangle$, $\langle\overline{P}\rangle$, $\langle
Q\rangle$, $\langle \overline{Q}\rangle \neq 0$ with the
inflationary superpotential given by~\cite{422},
\begin{eqnarray}
&&W_{\rm infl}\approx -\kappa S\bigg[ M_{B-L}^2- {\bf
16}_H{\bf\overline{16}}_H+\frac{\rho}{\kappa M_*^2 }({\bf
16}_H{\bf\overline{16}}_H)^2
\nonumber \\
&&-\frac{\kappa_1}{\kappa}P\overline{P}+\frac{\rho_1}{\kappa
M_*^2}(P\overline{P})^2
-\frac{\kappa_2}{\kappa}Q\overline{Q}+\frac{\rho_2}{\kappa M_*^2}
(Q\overline{Q})^2 \bigg] \nonumber
\\ && \equiv -\kappa SM_{\rm
eff}^2 ~,
\end{eqnarray}
where $M_{\rm eff}^2$ turns out to be of order $M_{B-L}^2$.  With
$D_SW\approx -\kappa M_{\rm eff}^2(1+|S|^2/M_P^2)$,
Eq.~(\ref{scalarpot}) becomes
\begin{eqnarray} \label{inflpot}
V_{F}&\approx\bigg(1+\sum_{k}\frac{|\phi_k|^2}{M_P^2}+\cdots\bigg)\bigg[
\kappa^2M_{\rm eff}^4\bigg(1+\frac{|S|^4}{2M_P^4}\bigg)
~~~~\nonumber \\
&~~~+\bigg(1+\frac{|S|^2}{M_P^2}+\frac{|S|^4}{2M_P^4}\bigg)
\sum_{k}|D_{\phi_k}W|^2\bigg] ~,
\end{eqnarray}
where all scalar fields except $S$ contribute to $\phi_k$.  In
Eq.~(\ref{inflpot}) the quadratic term of $S$ from $|D_SW|^2$,
which is of order $(\kappa^2M_{\rm eff}^4/M_P^2)|S|^2$ ($\approx
H^2|S|^2$), has canceled out with the factor ``$-3|W_{\rm infl
}|^2/M_P^2$'' and the quadratic term in $S$ from
``$e^{K/M_P^2}$.'' It is a common feature in this class of
models~\cite{review}.  Thus, the dominant mass term for $S$ is
\begin{eqnarray}\label{sugracor}
V_F\supset \sum_{l} |D_{\phi_l}W|^2 \sim
\bigg(\frac{M_{GUT}}{M_P}\bigg)^2\times H^2|S|^2 ~,
\end{eqnarray}
where $\phi_l=X^{(')}, Y, {\bf 45}_H, {\bf 45}_H'$, and $H$
($\approx \kappa M_{\rm eff}^2/M_P$) denotes the ``Hubble induced
mass.'' Such a small mass term of $S$ ($<<H^2|S|^2$) does not
spoil the slow roll conditions.  Note that the $U(1)$ $R$-symmetry
ensures the absence of $S^2$, $S^3$, etc. in the superpotential,
which otherwise could spoil the slow-roll conditions.


At one of the local minima, $\langle{\bf 16}_H\rangle$,
$\langle{\bf \overline{16}}_H\rangle$, $\langle P\rangle$,
$\langle\overline{P}\rangle$, $\langle Q\rangle$, and
$\langle\overline{Q}\rangle$ acquire the non-zero VEVs; $|\langle
{\bf 16}_H\rangle|^2=|\langle{\bf\overline{16}}_H\rangle|^2
\approx \kappa M_*^2/(2\rho)$, $|\langle
P\rangle|^2=|\langle\overline{P}\rangle|^2\approx \kappa_1
M_*^2/(2\rho_1)$, and $|\langle
Q\rangle|^2=|\langle\overline{Q}\rangle|^2\approx \kappa_2
M_*^2/(2\rho_2)$~\cite{422}.  Since $P$ and $Q$ develop VEVs,
$X^{(')}$, $Y$, ${\bf 45}_H$, and ${\bf 45}_H'$ should also
achieve VEVs from $D_{{\bf 16}^{(')}}W=D_{{\bf
\overline{16}}^{(')}}W=0$ even during inflation. Consequently,
$SO(10)$ and $U(1)_A$ are broken to $G_{SM}$ during inflation.
Note that $\langle P\rangle=\langle \overline{P}\rangle$ and
$\langle Q\rangle=\langle \overline{Q}\rangle$ lead to $\langle
P_-\rangle=\langle Q_- \rangle=0$. Since $\langle P_-\rangle$ and
$\langle Q_-\rangle$ vanish both during and after inflation,
oscillations by such light ($\sim m_{3/2}$) scalars would not
arise after inflation has ended. A non-zero vacuum energy from the
``F-term'' potential induces universal ``Hubble induced scalar
mass terms'' ($\approx\kappa^2 M_{\rm eff }^4/M_P^2\times
|\phi_l|^2$), which are read off from Eq.~(\ref{inflpot}). But
such small masses ($<<M_{B-L}$) can not much affect the VEVs of
the superheavy scalars of order $M_{GUT}$.


With SUSY broken during inflation ($F_S\neq 0$), there are
radiative corrections from the ${\bf 16}_H$,
${\bf\overline{16}}_H$ supermultiplets, which provide logarithmic
corrections to the tree level potential $V_F\approx\kappa^2M_{\rm
eff}^4\approx \kappa^2M_{B-L}^4/(4\zeta)^2$, and thereby drive
inflation~\cite{hybrid}. In our model, the scalar spectral index
turns out to be $n_s=0.99\pm 0.01$ for $\kappa<10^{-2}$, and the
symmetry breaking scale $M_{B-L}$ is estimated to be around $
10^{16}-10^{17}$ GeV~\cite{ss}.  When inflation is over, the
inflatons decay into right handed neutrinos. Following
Ref.~\cite{dec}, the lower bound on $T_r$ is $T_r\lapproxeq 10^9$
GeV for $\kappa\lapproxeq 10^{-2}$. The inflaton decay into right
handed neutrinos yields the observed baryon asymmetry via
leptogenesis~\cite{lepto-inf}. Assuming non-thermal leptogenesis
and hierarchical right handed neutrinos, we estimate the three
right handed neutrinos masses to be of order $10^{14}$ GeV,
$(10-20)\times~T_r$ and few $\times~ T_r$. Note that with $\kappa
< 10^{-2}$ the inflaton (with mass $\sim\sqrt{\kappa M_{B-L}^2}$)
can not decay into the heaviest right handed neutrino (of mass
$\sim 10^{14}$ GeV). Thus, the latter does not play a direct role
in leptogenesis.


In summary, our goal here was to realize inflation in a realistic
SUSY $SO(10)$ model. A global $U(1)_A$ and the $U(1)$ $R$-symmetry
plays essential roles in the analysis. The scalar spectral index
is $n_s=0.99\pm 0.01$, which will be tested by several ongoing
experiments. Proton decay proceeds via $e^+ \pi^0$, with an
estimated lifetime of order $10^{34}-10^{36}$ yrs. The LSP is
stable. While the heaviest right handed neutrino weighs around
$10^{14}$ GeV, the one primarily responsible for non-thermal
leptogenesis has mass of order 10 $T_r$, where the reheat
temperature $T_r$ is around $10^8-10^9$ GeV.
%
%
%
%

\end{document}